\documentclass[]{spie}  

\usepackage{amsmath,amsfonts,amssymb}
\usepackage{graphicx}
\usepackage[colorlinks=true, allcolors=blue]{hyperref}
\title{Modeling luminescent coupling in multi-junction solar cells: Perovskite Silicon tandem case study }

\author[a,b]{Phillip~Manley}
\author[a,b]{Martin~Hammerschmidt}
\author[a,b]{Lin~Zschiedrich}
\author[a,c]{Klaus~J\"{a}ger}
\author[c]{Christiane~Becker}
\author[a,b]{Sven~Burger}

\affil[a]{Computational Nano Optics, Zuse Institute Berlin, Takustraße 7, 14195 Berlin, Germany}
\affil[b]{JCMwave GmbH, Bolivarallee 22, 14050 Berlin, Germany}
\affil[c]{Department Optics for Solar Energy, Helmholtz-Zentrum Berlin für Materialien und Energie~GmbH, Albert-Einstein-Straße 16, Berlin 12489, Germany}

\authorinfo{E-mail: phillip.manley@jcmwave.com}

\begin{document} 

\maketitle

\begin{abstract}
Luminescent coupling is a characteristic of multi-junction solar cells which has often been neglected in models of their performance. The effect describes the absorption of light emitted from a higher band gap semiconductor by a lower band gap semiconductor. In this way, light which might have been lost can be utilized for current generation. We present a framework for modeling this effect in both planar layer stacks and devices with periodic nanostructuring. As a case study, we evaluate how luminescent coupling is affected by the inclusion of nanostructuring in a perovskite silicon tandem solar cell. We find that nanostructuring, while reducing the reflection loss for tandem solar cells also reduces the luminescent coupling, allowing more light to be emitted to the surroundings, when compared to planar devices. This highlights the need to include modeling of this effect into optimization schemes in order to find the trade-off between these two effects. The published version of this work is available at https://doi.org/10.1117/12.3023941.
\end{abstract}


\section{INTRODUCTION}
\label{sec:intro}  
In striving for the highest possible efficiency solar cells to help combat the climate crisis, one aspect that has yet to be fully exploited is the luminescent coupling of tandem, and more broadly multi-junction, devices. When any solar cell is operating at the maximum power point, a certain fraction of the electron hole pairs generated via absorption of solar radiation will recombine before being extracted. This recombination will be either radiative, i.e., emission of a photon, also called luminescence, or non-radiative, in which case heat is generated. The luminescence can be further broken down into light which is re-absorbed and light which is lost through emission to the exterior. If the re-absorption of the light occurs in a different absorber layer of the multi-junction device, the process is called luminescent coupling \cite{steiner2013measuring}.

Typical tandem and multi-junction systems employ absorbing materials of different band gap energies, with the higher energy band gaps placed towards the front of the device. This allows for those absorbers to filter out the high energy (short wavelength) sunlight and allow the lower energy (longer wavelength) light through to lower lying layers. Since luminescence occurs approximately at the band gap energy of an absorbing material, luminescent coupling is only possible when going from a higher band gap energy absorber to a lower band gap energy absorber. If the lower band gap energy absorber emits a photon, the energy of that photon will be too low to be absorberd by the higher band gap absorber. Therefore, in tandem configurations, with only two absorbers, luminescent coupling can only happen from the upper absorber to the lower absorber \cite{baur2007effects}. In the case of a perovskite-silicon tandem, used as a case study in this contribution, that means emission from the perovskite into the silicon. For higher numbers of junctions, the situation is more complicated, but the general rule that the luminescence can only be usefully coupling towards the back of the solar cell generally holds.

In the ideal case, a solar cell absorber exhibits purely radiative recombination and the re-emitted light is almost completely reabsorbed in the layer in which it was emitted. If the absorber is part of a multi-junction system, this means no luminescent coupling takes place. Both of these conditions can be approached in single junction GaAs solar cells \cite{miller2012strong, balenzategui2006detailed}. This leads to a build of the photon density inside the absorbing layer which increases the open circuit voltage of the device \cite{ganapati2016voltage}. Note that in order to achieve this, light that is emitted in the absorbing layer has to be effectively trapped. For a planar GaAs absorber with a highly reflective mirror on the rear side, light can be very efficiently trapped via total internal reflection at the front surface \cite{gruginskie2021limiting}. However, once the device consists of multiple layers, it becomes difficult to trap the luminescence of one absorbing layer while letting incident sunlight through to be absorbed in lower lying absorbing layers. This means that multi-junction devices do not typically benefit significantly from the increased open circuit voltage due to re-absorption of luminescence \cite{sheng2015device}. This can be partially compensated for via luminescent coupling, which leads to an increased current in lower lying absorber layers.

Luminescent coupling has the further beneficial property of relaxing the conditions for high efficiency operation of the tandem solar cell. Typically, the two cells in a tandem device are designed to be monolithic and series connected in order to reduce losses. This imposes the condition that the two cells be current matched, i.e., the same current must flow through both cells. Therefore, whichever cell has the lower current will limit the overall device efficiency. Tandem cells are designed such that the same current flows through each sub-cell under ideal conditions. If these conditions change, due to the movement of the sun, weather or shading effects, then the current distribution will no longer be matched, limiting the overall efficiency. If the current is higher in the top cell than the bottom cell, this leads to an increase in the luminescent coupling to the bottom cell, thereby partially equalizing the two currents \cite{Bowman_2021,tillmann2021relaxed}. This means that the tandem device can maintain high efficiencies over a larger range of conditions.


To obtain the highest efficiency devices, it is necessary to include modeling of both the luminescence re-absorption and luminescent coupling. At the same time, these two effects should be balanced against the need to get sunlight into the device (anti-reflection) and ensure that as much of the incident sunlight is absorbed as possible. Typically, there will be a trade-off between these effects, warranting careful study to find the optimal configuration. 

One of the most promising device architectures for tandem solar cells is the combination of a thin film perovskite top cell and a crystalline silicon wafer bottom cell, which currently has achieved record efficiencies of 33.9\% \cite{green2024solar}. Therefore we will take this device as a case study for investigating the impact of luminescent coupling in tandem devices. The efficiency of luminescent coupling in this device architecture has been assessed for planar structures \cite{jager2021perovskite} with ray optical models. Here we seek to use a more realistic point dipole model to describe the internal emission, as well as to extend the model to periodic nanostructures.

This contribution is structured as follows, first the physical model and the numerical method used to find solutions to the model are described. This is followed by results and discussion for both the planar and nano\-structured devices. Finally the results of this work are summarised and avenues for improvement are discussed in the conclusion.

\section{NUMERICAL METHOD}
\begin{figure}
    \centering
    \includegraphics[width=0.6\textwidth]{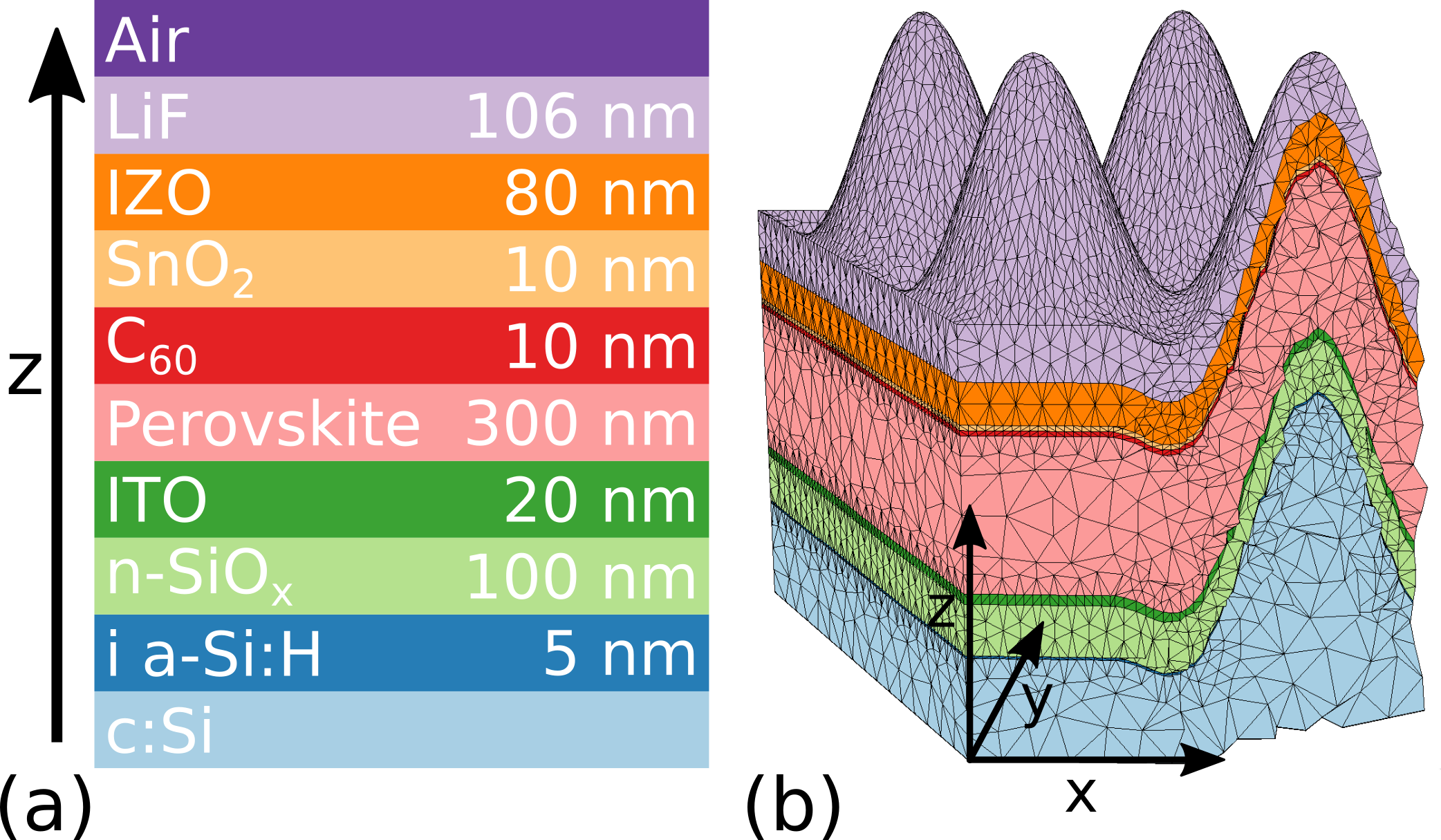}
    \caption{(a) The materials and their thicknesses in the planar perovskite-Si tandem solar cell. Note that the silicon is treated as a semi-infinite layer. (b) Cut through of a 2x2 supercell of the periodic nanostructure.}
    \label{fig:layers_and_mesh}
\end{figure}

The field of perovskite-Si tandem solar cells is rapidly advancing. This results in a multitude of different perovskite compositions, transport layer materials and different structure geometries present in the literature. Here we seek to present a method for evaluating the luminescent coupling for multi-junction systems in general. Therefore, we are choosing a multi-layer stack system similar to one that has been previously investigated for both luminescent coupling of planar layers and the reflection reduction due to nanostructuring \cite{jager2021perovskite, chen2018nanophotonic}. The stack system, shown in fig. \ref{fig:layers_and_mesh}(a) consists of (from top to bottom), a LiF anti-reflective coating, a two layer electric contact consisting of indium doped zinc oxide (IZO) and tin oxide, an electron transport layer (ETL) of pristine fullerene (C$_{60}$), the methylammonium lead iodide perovskite absorber layer (MAPbI$_{3}$), a second electrical contact in the form of indium doped SnO$_{2}$ (ITO), an isolated layer of n-doped nanocrystalline silicon oxide (nc-SiO$_{x}$), an ETL of intrinsic hydrogenated amorpous Si (i a-Si:H) and the crystalline silicon absorber layer (c:Si) which is here treated as semi-infinite. The associated layer thicknesses are present in fig. \ref{fig:layers_and_mesh}(a). Note that a hole transport layer (HTL) for the perovskite upper cell has been neglected due to a lack of readily available refractive index data. We do not expect that this layer will strongly affect the optical response of the device. 

The nanostructure consists of a sinusoidal texture on a hexagon lattice with a 500\,nm period and 500\,nm height. Further details to the nanostructure can be found here \cite{chen2018nanophotonic}.

To estimate the amount of luminescent coupling in a perovskite-Si tandem solar cell device, we model the emission of light inside the perovskite layer using a dipole source. Since we are interested in the steady-state response for continuous generation of light inside the perovskite (due to absorption of a continuous source of sunlight) it is appropriate to employ the time harmonic ansatz. The luminescence inside the perovskite will be narrowband in comparison to the incident solar spectrum. We have chosen to investigate the emission at a single wavelength of 800\,nm, close to the band edge of methylammonium lead iodide (MAPbI$_{3}$) perovskite. We note that a weighted spectral average using the experimentally derived emission profile of the perovskite material should be included in the model for a more robust analysis. 

The absorption of the emission in the various layers is obtained via volume integration of the electric field energy density, while the emission into the upper and lower semi-infinite spaces is obtained via a surface integral of the Poynting flux. Note that the total emission of a dipole in an absorbing medium in the time-harmonic regime is undefined. In order to have a well defined emission energy, the absorption in the entire perovskite layer was artificially set to zero. This will neglect the reabsorption of emission inside the perovskite and the associated increase in open circuit voltage. However, if the effect of non-radiative recombination is neglected, then despite undergoing many emission and reabsorption events inside the perovskite layer, all of the light must eventually make its way outside of the perovskite layer and contribute to either parasitic absorption in other layers, or to emission into the upper or lower domains. With the additional assumption that charge carrier diffusion lengths inside the absorber layer are larger than the layer thickness, we can assume that emission is homogeneous with respect to position inside the absorbing layer. Therefore, while the proposed model cannot estimate the increased open circuit voltage due to luminescence, the luminescent coupling will be accurately estimated.

This contribution investigates two different geometries, firstly a layered system with planar layers and secondly a layered system with a periodic nanostructure. Both of these geometries are presented in fig. \ref{fig:layers_and_mesh}. Due to the rotational symmetry present for the planar system with a dipole source positioned on the rotational axis, the light scattering problem can be converted from a three dimensional problem to a series of two dimensional problems via an expansion into Fourier modes~\cite{Schneider2018oe}. This greatly reduces the computational effort required.

The periodic nanostructure allows for the simulation of a single unit cell via the application of Floquet boundary conditions in the $x-y$ plane. This will have the consequence of including a dipole in each the infinite number of unit cells implied by the periodic boundary conditions. Furthermore the dipole sources will emit light coherently with one another. This is unphysical as the emission due to reabsorption is a stochastic process, meaning there should be no coherence relationship between emission at different positions in the perovskite. One solution to this problem is to simply increase the domain size to include more than one unit-cell of the periodic lattice. Such supercells will, in the limit of large sizes, approximate an isolated dipole emitting in a periodic environment. This will greatly increase the computational costs of the simulation. A different solution is to integrate over different phase jumps 
 applied at the Floquet boundaries \cite{zschiedrich2013numerical}. These phase jumps characterize different waves of different $k$ vectors travelling through the periodic lattice, and are commonly referred to as Bloch vectors. The integral can be done using separate single unit cell computations, which is generally more computationally efficient than a single large supercell computation. It also has the benefit of allowing for adaptive schemes for the integration, further reducing the computational cost.

The boundaries to air and the crystalline silicon substrate are modeled as transparent boundary conditions for both of the geometries. The transparent boundaries are achieved numerically using perfectly matched layers. 

In order to solve Maxwell's equations we employ the finite element solver JCMsuite \cite{pomplun2007adaptive}. The element sizes were chosen to be no larger than half the in-material wavelength, using polynomial orders between two and three, chosen locally on each element to obtain a goal accuracy of under 1\%. Refractive index data for the perovskite \cite{guerra2017determination} and crystalline Si \cite{green1995optical} were taken from literature. The refractive indices of all other layers are based on in-house ellipsometry measurements. 

For the planar layer stack, no reciprocal space integration is necessary
due to the cylindrically symmetric setup with a single, isolated point source. 
For the nanostructured system, an integral in reciprocal space is required. The reciprocal space integration was achieved by a discrete approximation employing a total of 102 simulations for different Bloch vectors distributed on a regular grid in the irreducible Brillouin zone. This corresponds approximately to a $16\times16$ supercell. 

For the real space integration in the planar device, 100 dipole sources equally spaced along the $z$ axis inside the perovskite layer were used. For the real space integration in the nanostructured device over source positions 100 different dipole source positions were simulated, each for three linearly independent polarizations. In order to obtain weighting factors for the volume integration, a two step process was employed. First the finite element mesh is used to produce a point cloud consisting of the centroids of each element in the perovskite domain, weighted by the respective volume of each element. Using this point cloud directly would necessitate simulating thousands of dipole source positions, which is computationally expensive. A k-means clustering algorithm is applied to the dense point cloud and their respective weightings to obtain a less dense point cloud with a desired number of points. The total weighting of the point clouds is conserved, meaning that the sum of weightings in the less dense point cloud also corresponds to the volume of the structured perovskite layer. The point cloud generated is shown in fig. \ref{fig:vol_kspace_integration}.


\section{RESULTS \& DISCUSSION}

\begin{figure}
    \centering
    \includegraphics[width=\textwidth]{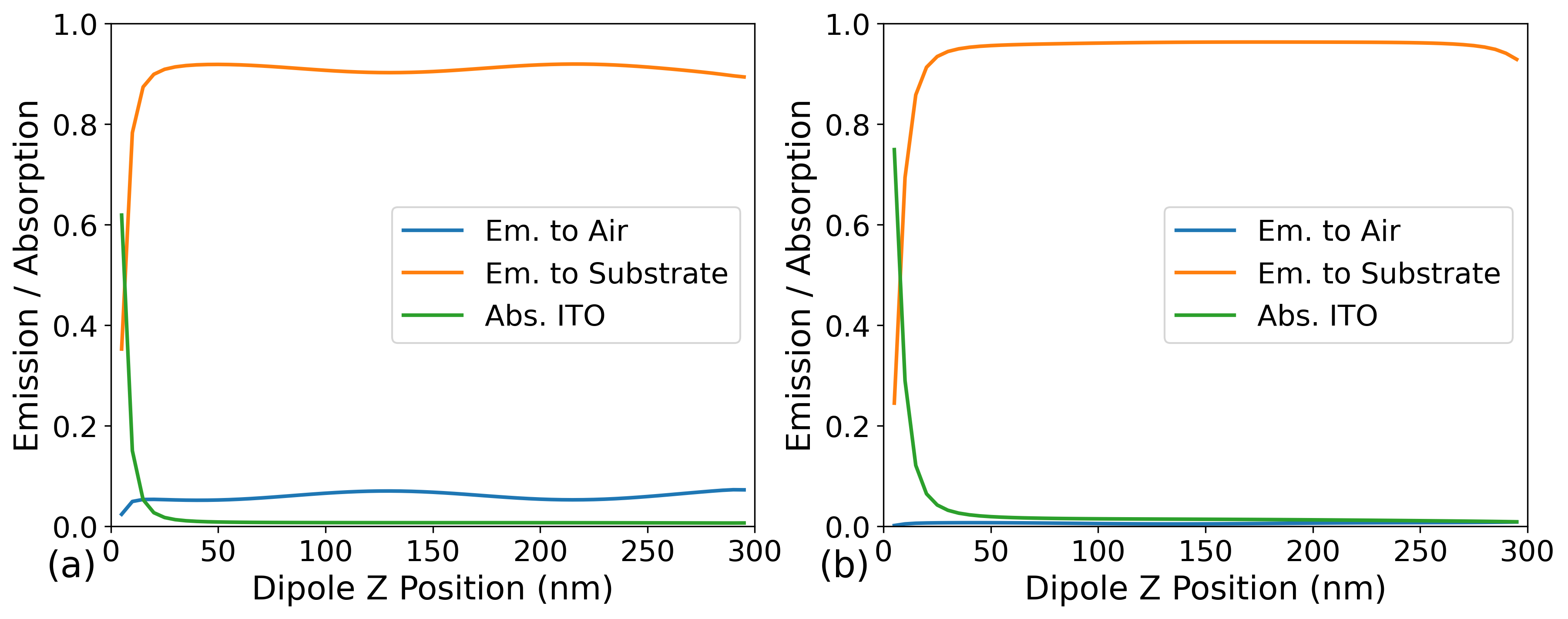}
    \caption{Planar layer stack: Dependence of emission and absorption loss on dipole position in the perovskite layer for (a)\,in-plane polarized dipoles and (b)\,$z$ polarized dipoles. Note that the perovskite is modeled as a non-absorbing layer.}
    \label{fig:planar_positional_emission}
\end{figure}

Figure \ref{fig:planar_positional_emission} shows the emission dependence on the $z$ position of a dipoles in the planar perovskite layer as part of the tandem system shown in fig. \ref{fig:layers_and_mesh}(a). Due to the rotational symmetry, it is sufficient to look at only one in-plane polarization (a), along with the out of plane, i.e., $z$, polarization (b). The in-plane polarized dipoles show a very slight interference effect depending on their position in the $z$ direction. This effect is small since most of the emission is efficiently able to couple out of the perovskite layer into the underlying Si layer. If the dipoles become very close to the interface to the ITO layer which is directly beneath the perovskite, evanescent coupling into the thin ITO layer dominates and the absorption into the ITO rapidly increases at the cost of emission into the substrate. The situation for $z$ polarized dipoles is similar, though interference effects are not present. The quenching of emission via absorption in the ITO is more pronounced. This is likely due to easier coupling to evanescent modes of the ITO for $z$ polarized dipoles. As previously mentioned, the model does not include a HTL between the perovskite and ITO contact. Inclusion of a HTL would show quenching of the emission in the ETL itself, if the ETL is highly absorbing, or in the ITO if the ETL is transparent at the emission wavelength of 800\,nm. Note that the spatial distribution would be of emission would be affected when including absorption in the perovskite layer, with lower emission to the substrate from the dipoles situated at the front side of the perovskite.

It should be noted that for planar systems with effective light barriers on both sides, for example a mirror below and an air interface above, the difference between polarizations would be much more pronounced. In that case, light from $z$ polarized dipoles can be effectively trapped via total internal reflection. This is often the case for LED structures. Here, since the Si substrate is also a high index dielectric material, there is no total internal reflection barrier.

When integrating over the different $z$ positions and polarizations, a total emission of 92\% to the Si substrate is obtained for isotropically distributed dipoles. This is in agreement with a previous estimate of 91\%, if the re-absorption in perovskite is neglected and the resulting emission rates are renormalized \cite{jager2021perovskite}.

\begin{figure}
    \centering
    \includegraphics[width=\textwidth]{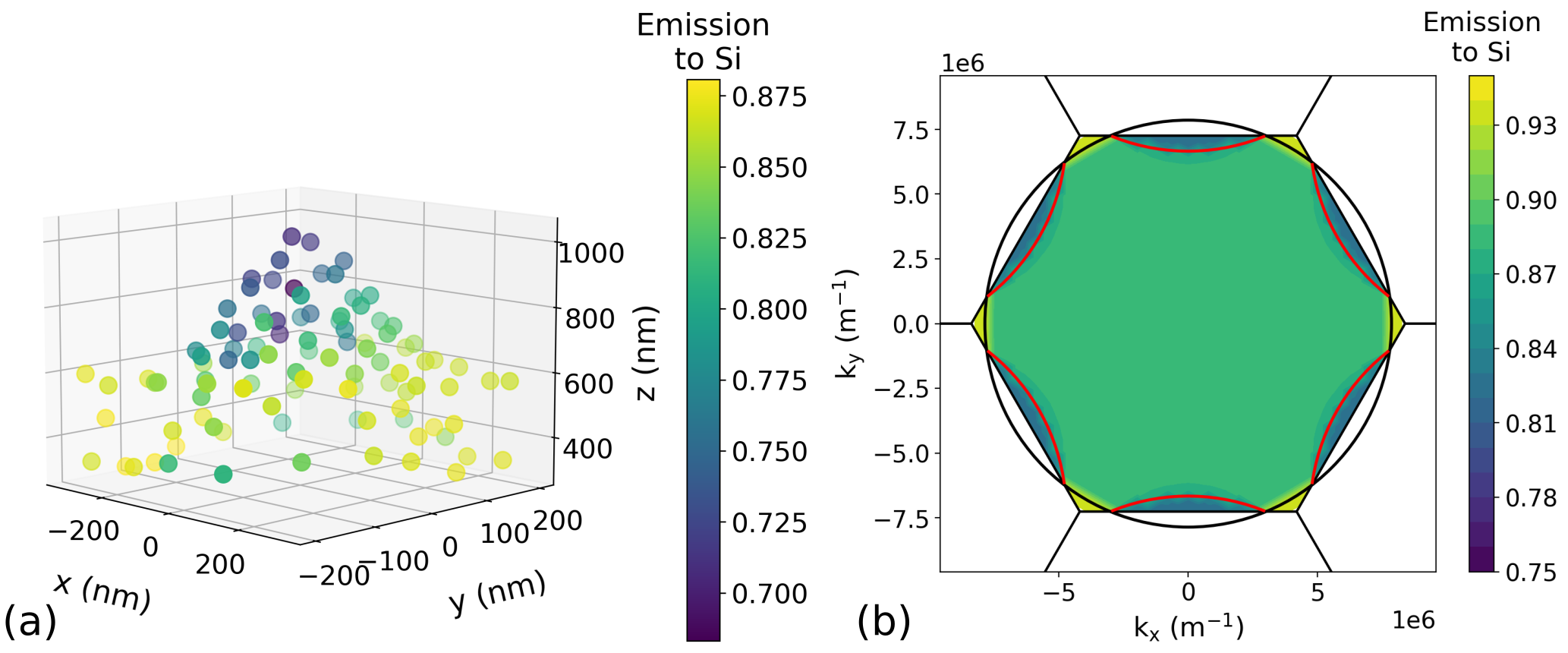}
    \caption{(a) The emission into the Si substrate for the different dipole positions used for the volume integration of the nanostructured tandem device. (b) The emission into the substrate distributed in the first Brillouin zone for the nanostructured tandem device. The hexagonal reciprocal lattice is indicated, as well as the propagating spectrum in air (black circle). The red lines indicate the positions of the disappearance of the first diffraction order in air.}
    \label{fig:vol_kspace_integration}
\end{figure}

Next, we turn our attention to the nanostructured device. To obtain an estimate for the emission into the Si substrate for isotropically distributed dipoles, each emitting incoherently with each other, it was necessary to perform the integrations over real and reciprocal space as described in the methods section. Figure \ref{fig:vol_kspace_integration}(a) shows the distribution of the emission into the substrate over the volume of the nanostructured perovskite layer. The emission to the substrate is high for the dipoles positioned in the valleys of the nanostructure. Conversely, the emission to the substrate drops from above 85\% to around 70\% for dipoles situated at the tips of the nanostructures. This can be explained by the fact that those dipoles are effectively surrounded by an air interface. This will suppress reflection as light emitted from the dipole reaches the air interface more or less at normal incidence. Note that the emission at each position has been obtained via an integration over the Bloch phase in reciprocal space.

The distribution of emission into the substrate can also be viewed in reciprocal space. Figure \ref{fig:vol_kspace_integration}(b) shows, analogously to (a), the distribution of emission into the substrate. Here each position in reciprocal space is the average obtained via integrating over all positions in real space. For the majority of k-vectors, the emission into the substrate is around 90\%. At the edges of the Brillouin zone we see rapid changes in the emission. For light emission with transverse $k$ vectors outside the propagating spectrum in air, i.e., the parts of the first Brillouin zone lying outside the black circle, emission into air is impossible and the emission into the Si substrate is greatly enhanced. On the other hand, when additional diffraction orders for light emission into air become available inside the first Brillouin zone (indicated by the red lines) the emission into the substrate is greatly reduced as light has additional pathways to escape the device through the higher diffraction orders in air. This indicates that the emission to the substrate could be enhanced by employing nanostructures with smaller periods, thereby increasing the size of the Brillouin zone with respect to the propagating spectrum in air. This would lead to more light emission into regions where propagation in air is  not possible and would push the diffraction orders out of the first Brillouin zone.

When averaging over both real and reciprocal space, an estimate for the emission into the Si substrate of  83\% is obtained for isotropically distributed dipoles in the nanostructured perovskite layer. This is a reduction compared to the planar case with 92\%. The nanostructure was initially introduced to reduce reflection losses for incident sunlight. A structure which is able to more efficiently couple light into the device will, due to reciprocity, also more easily couple light out of the device. Therefore, it is expected that the emission to the substrate will decrease via the inclusion of the nanostructure. In order to fully optimize the optical properties of this and other multi-junction devices, it is critical to estimate both the positive and negative optical effects of introducing nanostructured layers. With accurate models a trade-off can be found in order to obtain the lowest reflection losses while maintaining high luminescent coupling efficiency. 
	
\section{CONCLUSION}
We have presented a model for estimating the amount of luminescent coupling in nanostructured multi-junction solar cells. This model employs integration over both real and reciprocal space to approximate the response of an isotropically distributed ensemble of dipoles emitting light incoherently with each other in a periodically structured environment.

It was shown that for the planar case, 92\% of luminescence in the perovskite layer will be coupled to the underlying Si substrate, in line with other estimates in the literature. The nanostructured cell was shown to reduce the luminescent coupling to 83\%. This emphasises the need to model luminescence as part of a complete optical optimization for solar cell devices.

Future work should seek to include the re-absorption of luminescence into the model to estimate the increase to the open circuit voltage this provides. Furthermore, the finite bandwidth of luminescence should also be taken into account. Non-radiative recombination effects should also be included to accurately estimate the potential of luminescence for solar cell devices.

\acknowledgments 
 
The results were obtained at the Berlin Joint Lab for Optical Simulations for Energy Research (BerOSE) of Helmholtz-Zentrum Berlin für Materialien und Energie GmbH, Zuse Institute Berlin and Freie Universität Berlin.
This project has received funding from the German Federal Ministry of Education and Research (BMBF Forschungscampus  MODAL, project number 05M20ZBM).
The authors would like to thank Steve Albrecht, Alvaro Tejada Esteves and Florian Ruske for providing refractive index data.

\bibliography{main} 
\bibliographystyle{spiebib} 

\end{document}